\documentclass{article}

\usepackage{microtype}
\usepackage{graphicx}
\usepackage{subfigure}
\usepackage{booktabs} 

\usepackage{hyperref}

\usepackage[accepted]{icml2025}

\usepackage{amsmath}
\usepackage{amssymb}
\usepackage{mathtools}
\usepackage{amsthm}
\usepackage[capitalize,noabbrev]{cleveref}
\usepackage{lipsum}
\usepackage{stfloats} 
\theoremstyle{plain}

\theoremstyle{definition}

\theoremstyle{remark}

\usepackage{makecell}
\usepackage{mathrsfs}
\usepackage{algorithm}
\usepackage{algorithmic}
\usepackage{multirow}
\usepackage{graphicx}

\usepackage[textsize=tiny]{todonotes}

\icmltitlerunning{}

\begin{document}

\twocolumn[
\icmltitle{Topology of surface electromyogram signals: hand
gesture decoding on Riemannian manifolds}



\icmlsetsymbol{equal}{*}

\begin{icmlauthorlist}
\icmlauthor{Harshavardhana T. Gowda}{yyy}
\icmlauthor{Lee M. Miller}{zzz,ccc,vvv,y}
\end{icmlauthorlist}

\icmlaffiliation{yyy}{Department of Electrical and Computer Engineering, University of California, Davis}
\icmlaffiliation{zzz}{Center for Brain and Mind, University of California, Davis}
\icmlaffiliation{ccc}{Department of Neurobiology, Physiology, and Behavior, University of California, Davis}
\icmlaffiliation{vvv}{Department of Otolaryngology/Head and Neck Surgery, University of California, Davis}
\icmlaffiliation{y}{\textcolor{blue}{A version of this article has been published in the Journal of Neural Engineering. {\em J. Neural Eng. 21 036047}}}

\icmlcorrespondingauthor{Harshavardhana T. Gowda}{tgharshavardhana@gmail.com}

\icmlkeywords{Machine Learning, ICML}

\vskip 0.3in
]



\printAffiliationsAndNotice{}  

\begin{abstract}

{\em Objective.}
In this article, we present data and methods for decoding hand gestures using surface electromyogram (EMG) signals. EMG-based upper limb interfaces are valuable for amputee rehabilitation, artificial supernumerary limb augmentation, gestural control of computers, and virtual and augmented reality applications.
{\em Approach.}
To achieve this, we collect EMG signals from the upper limb using surface electrodes placed at key muscle sites involved in hand movements. Additionally, we design and evaluate efficient models for decoding EMG signals.
{\em Main results.}
Our findings reveal that the manifold of symmetric positive definite (SPD) matrices serves as an effective embedding space for EMG signals. Moreover, for the first time, we quantify the distribution shift of these signals across individuals.
{\em Significance.}
Overall, our approach demonstrates significant potential for developing efficient and interpretable methods for decoding EMG signals. This is particularly important as we move toward the broader adoption of EMG-based wrist interfaces.
\end{abstract}

\textbf{Data and code availability}\\
Our dataset is publicly available through the following link: 
\href{https://doi.org/10.17605/OSF.IO/ZCR43}{DOI: 10.17605/OSF.IO/ZCR43}.\\
Codes are available on GitHub: \href{https://github.com/HarshavardhanaTG/GeometryOFsEMG}{GitHub}.

\newpage
\section{Introduction}
\label{Introduction}
EMG signals are recorded non-invasively by placing sensors on the skin surface to measure electrical activity generated by motor unit activation. The global characteristics of the EMG signal, such as amplitude and power spectrum, depend on numerous idiosyncratic factors. These include anatomical characteristics, such as the thickness of subcutaneous tissue, the distribution and size of motor unit (MU) territories, and the spread of endplates and tendon junctions within the MU. Physiological factors, such as the distribution of conduction velocities within motor unit fibers, the shape of intracellular action potentials \cite{farina2004extraction}, and muscle fatigue \cite{enoka2008muscle}, also influence the signal. Additionally, circumstantial factors, such as precise electrode placement \cite{HUEBNER2015214, kleine2001influence}, further contribute to signal variability. The combined effects of these factors are further complicated by interactions between signals originating from multiple neighboring muscles. As a result, signals from individual EMG electrodes tend to be highly confounded and opaque, limiting their practical utility.

We demonstrate that covariance matrices, constructed using the pairwise covariance of electrical signals measured across different electrodes, effectively capture the combined influence of various physiological and anatomical factors. Consequently, this approach provides a robust framework for quantifying differences in EMG signals across individuals. Furthermore, the spatial patterns captured by these covariance matrices exhibit rich geometric structures, which can be effectively leveraged to distinguish different hand gestures.

\subsection{Prior work}
\noindent Existing methods rely on constructed features such as signal root-mean-square, time-domain statistics, as described by \citeauthor{204774}, histograms \cite{481972}, marginalized discrete wavelet transform \cite{LUCAS2008169}, or a normalized combination of all the above. These features are typically evaluated using classifiers such as linear discriminant analysis (LDA), support vector machines (SVM), {\em k}-nearest neighbors ({\em k}-NN), and random forests.

Additionally, some studies employ deep learning models, including convolutional neural networks (CNN) \cite{9054586, geng2016gesture, wei2019multi, khushaba2022myoelectric}, recurrent neural networks (RNN) \cite{8333395, 9463745}, transformer-based networks \cite{rahimian2021temgnet, Montazerin2023}, and hybrid architectures combining RNN and CNN \cite{rahimian2020surface, hu2018novel}. \citeauthor{xiong2022learning} analyze EMG signals using SPD covariance matrices; however, by mapping the learned features onto a tangent plane and decoding them in Euclidean space, this approach fails to leverage the natural geometric structure of the data.

Despite their advancements, all these methods overlook the strong spatial correlations in muscle contraction patterns. Furthermore, they often involve a large number of trainable parameters - ranging in the tens of thousands - and require complex transfer learning paradigms and extensive retraining for deployment across individuals. Moreover, none of these established techniques readily adapt to signal variations caused by factors such as muscle fatigue and sensor displacement.

\subsection{Our contribution}
To overcome the limitations of existing methods, we propose a more principled approach by analyzing EMG signals on a Riemannian manifold, which offers a more comprehensive and natural framework for representing the data structure compared to conventional Euclidean analysis. Specifically, the SPD covariance matrices derived from multivariate EMG time series form a structured representation on a cone manifold with a Riemannian metric. This geometric framework naturally preserves the spatial correlations in muscle contraction patterns.

To ensure computational efficiency and numerical stability, we study these SPD matrices via Cholesky decomposition in the Cholesky space \cite{Lin_2019}, which consists of lower triangular matrices with positive diagonal elements. We demonstrate that covariance matrices from different individuals occupy distinct neighborhoods within the manifold space due to the combined influence of anatomical and physiological factors on EMG signals. Furthermore, we quantify inter-individual differences in EMG signals using the geodesic distance - the length of the shortest curve between two points on a surface (between corresponding covariance matrices).

Leveraging this geometric structure, we introduce two supervised learning algorithms - manifold minimum distance to mean and manifold support vector machine - along with one unsupervised learning algorithm, manifold {\em k}-medoids clustering, to classify hand gestures directly on the Riemannian manifold. This approach not only mitigates the challenges associated with existing feature extraction techniques and deep learning models but also offers a more interpretable and data-efficient alternative for EMG-based gesture recognition.

\section{Methods and materials}
We use three different datasets to demonstrate the efficacy of our methods. First, we utilize the widely used \textsc{Ninapro} (Non-Invasive Adaptive Prosthetics) \textsc{Database 2-Exercise 1}, provided by \citeauthor{atzori2014electromyography} Second, we employ a high-density EMG dataset from \citeauthor{Malešević2021} Third, we use a dataset collected by us, named \textsc{UCD-MyoVerse-Hand-0}. The first two datasets allow us to compare our methods with previously benchmarked approaches, while the inclusion of our own dataset ensures additional validation.

By selecting three distinct datasets, we aim to verify that our methods perform consistently well across different data collection platforms and experimental setups. Below, we provide a detailed explanation of the data acquisition protocols and EMG processing methods.

\subsection{EMG data acquisition setup}
A total of 30 subjects participated in our study (\textsc{UCD-MyoVerse-Hand-0}). Please refer to the {\em ethical statement} for the subject selection criteria. 

For \textsc{UCD-MyoVerse-Hand-0}, we used Delsys double differential EMG electrodes (\href{www.delsys.com}{Delsys, Inc.}) and an NI USB-6210 multifunction I/O device (\href{www.ni.com}{National Instruments Corporation} – 16 inputs, 16-bit, 250 kS/s) to acquire EMG data at a sampling rate of 2000 Hz. The Delsys electrodes transmitted the acquired data wirelessly to a base station, which then relayed the data to a computer via a USB connection through the NI USB-6210 data acquisition system.

A graphical user interface (GUI) was designed to display hand gestures on a screen. Subjects performed the displayed gesture with their dominant hand while seated comfortably in a chair, with their forearm resting on an elevated platform on the table. They were allowed to choose a resting position that was most comfortable for them and could adjust it throughout the experiment. Each gesture was performed for 2 {\em s}, followed by a resting period of 2 {\em s}. Specifically, the gesture was displayed on the screen for 2 {\em s}, followed by a blank screen for another 2 {\em s}. Subjects were instructed to perform the gesture for the duration the image was visible and to rest during the blank screen period.

The experiment was divided into six sessions, each consisting of sixty trials - six repetitions for each of the ten gestures. The order of gestures within a session was pseudorandomly generated to assess how variations in performing gestures affect decoding accuracy. If all six repetitions of a gesture occurred sequentially, it could lead to repetitive, unconscious, and highly consistent movements, which we aimed to avoid. In total, each subject completed 360 trials.

To synchronize GUI instructions with data streamed from the Delsys system, we used ZeroMQ sockets (\href{https://zeromq.org/socket-api/}{ZeroMQ}) and Lab Streaming Layer (\href{https://github.com/sccn/labstreaminglayer}{LSL}) in Python. Data streams were synchronized to the master clock on the computer, which received both EMG data and event markers from the GUI. We observed a maximum clock drift of approximately 50 {\em ms}.

\subsubsection{EMG of different hand gestures in UCD-MyoVerse-Hand-0}
Forearm EMG was collected from intact subjects using twelve electrodes. Eight electrodes were placed evenly spaced around the main belly of the forearm muscles, positioned below the elbow at approximately one-third of the distance from the elbow to the wrist. The remaining four electrodes were placed evenly spaced around the wrist joint.

\begin{figure}[h!]
	\centering
	\includegraphics[width=0.4\textwidth, angle = -90]{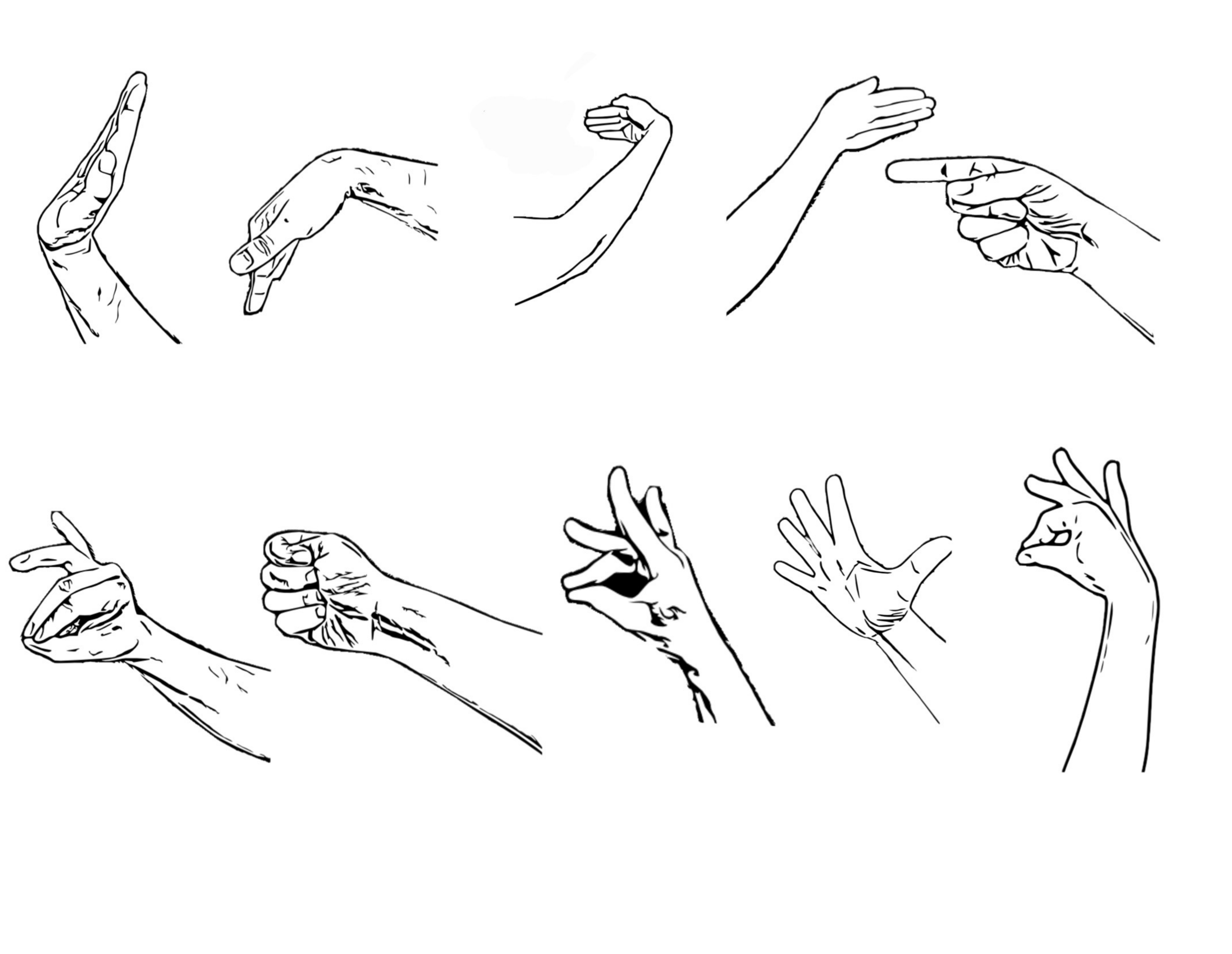}
	\caption{Ten gestures included in the \textsc{UCD-MyoVerse-Hand-0} experiment. From top-left: up, down, left, right, index point, two finger pinch, power grasp, middle finger pinch, splay, index finger pinch.}
	\label{fig:Gestures}
\end{figure}

Each subject performed ten different hand gestures, with each gesture repeated thirty-six times. The ten gestures included wrist movements in four cardinal directions (up, down, left, and right), three types of pinches (index finger pinch, middle finger pinch, and two-finger pinch), splay, power grasp, and pointing index (figure \ref{fig:Gestures}). These gestures were selected to reflect commonly used interactions with computers. Wrist movements in cardinal directions can be used for navigating the screen (moving it up, down, left, and right), while the pinch gestures can be used for zooming in, zooming out, and making selections.

\subsubsection{EMG of different hand gestures in \textsc{Nianpro: Database 2-Exercise 1}}
The dataset consists of EMG signals collected from forty intact subjects using twelve electrodes at 2000 Hz. Eight electrodes were placed evenly spaced around the forearm at the height of the radio-humeral joint, while two electrodes were placed on the main activity spots of the flexor digitorum superficialis and the extensor digitorum superficialis.

Subjects performed seventeen different gestures, each repeated six times. The seventeen gestures include: thumbs up, extension of the index and middle fingers with flexion of the others, flexion of the ring and little fingers with extension of the others, thumb opposing the base of the little finger, abduction of all fingers, fingers flexed together in a fist, pointing index, adduction of extended fingers, wrist supination (axis: middle finger), wrist pronation (axis: middle finger), wrist supination (axis: little finger), wrist pronation (axis: little finger), wrist flexion, wrist extension, wrist radial deviation, wrist ulnar deviation, and wrist extension with a closed hand.

For further details on data acquisition and processing, see \citeauthor{atzori2014electromyography}

\subsubsection{high density EMG of different hand gestures in \citeauthor{Malešević2021}}
The dataset consists of EMG signals collected from nineteen intact subjects using 128 electrodes placed at the forearm level at 2048 Hz. Although the experiment initially included twenty subjects, the data for {\em subject 5} was corrupted and therefore excluded.

Subjects performed 65 unique gestures, each formed by combining 16 basic single-degree-of-freedom movements. Each gesture was repeated five times.

For further details on data acquisition and processing, refer to \citeauthor{Malešević2021}

\subsection{EMG data preprocessing}
The data collection environment was carefully controlled to eliminate AC electrical interference. EMG signals undergo minimal preprocessing. Signals are $z$-normalized along the time dimension for each channel. The preprocessed EMG signals are then used to construct covariance SPD matrices. We detail the EMG analysis methods in the following section.
\subsection{EMG data analysis}
EMG signals are collected using a set of sensors, denoted as $\mathcal{V}$, and are represented as a matrix $\mathcal{X}$ with dimensions $(\mathcal{V}, \textsc{sampling frequency} \times \textsc{gesture duration})$. 

Signal covariance matrices are then computed from the preprocessed EMG signals using 
\[
\mathcal{E} = \mathcal{X} \mathcal{X}^T.
\]
The resulting matrices, $\mathcal{E}$, are symmetric and positive semi-definite. To ensure positive definiteness, we apply a shrinkage estimator as follows:  

\[
\mathcal{E} \leftarrow (1 - \eta)\mathcal{E} + \eta \, \texttt{trace}(\mathcal{E})\mathcal{I}
\]

where $\mathcal{I}$ is an identity matrix of the same dimensions as $\mathcal{E}$. Through empirical evaluation, we found that setting $\eta = 0.1$ is sufficient for all EMG data. The covariance matrices, $\mathcal{E}$, have dimensions $\mathcal{V} \times \mathcal{V}$.

Directly working with SPD matrices using affine-invariant or log-Euclidean metrics \cite{doi:10.1137/050637996} involves computationally expensive operations, such as matrix exponential and matrix logarithm calculations. These operations make mappings between the manifold space and the tangent space, and vice versa, computationally intensive.
To address this, \citeauthor{Lin_2019} proposed methods to operate on SPD matrices using Cholesky decomposition. They established a diffeomorphism between the Riemannian manifold of SPD matrices and Cholesky space. In Cholesky space, the computational burden is significantly reduced: logarithmic and exponential computations are restricted to the diagonal elements of the matrix, making them element-wise operations. Additionally, the Fréchet mean (centroid) is derived in a closed form. For an SPD edge matrix $\mathcal{E}$, the corresponding Cholesky decomposition $\mathcal{L} = \textsc{cholesky}(\mathcal{E})$ is such that $\mathcal{E} = \mathcal{L}\mathcal{L}^T$. A matrix $\lfloor\mathcal{L}\rfloor$ is the strictly lower triangular part of the matrix $\mathcal{L}$, and a matrix $\mathbb{D}(\mathcal{L})$ is the diagonal part of the matrix $\mathcal{L}$.
In the following section, we explain in detail, the methods used to analyze SPD matrices. 

\subsubsection{Distance between two SPD matrices $\mathcal{E}_1$ and $\mathcal{E}_2$}
The geodesic distance between two SPD matrices $\mathcal{E}_1$ and $\mathcal{E}_2$ is same as the distance between the corresponding Cholesky matrices $\mathcal{L}_1$ and $\mathcal{L}_2$ and is calculated as 
\begin{align}
    d(\mathcal{L}_1, \mathcal{L}_2) &= \left\{||\lfloor \mathcal{L}_1 \rfloor - \lfloor \mathcal{L}_2\rfloor||_F^2 \right. \nonumber \\
    &\quad + \left. ||\log\mathbb{D}(\mathcal{L}_1) - \log\mathbb{D}(\mathcal{L}_2)||_F^2\right\}^{1/2}, \label{eq:distance}
\end{align}
where $||.||_F$ denotes the Frobenius norm. 
\subsubsection{Fr\'echet mean (centroid) of SPD matrices}

Given a set of ({\em n}) SPD edge matrices $\mathcal{E}$, we first calculate their corresponding Cholesky decompositions $\mathcal{L} = \textsc{cholesky}(\mathcal{E})$, such that $\mathcal{E} = \mathcal{L}\mathcal{L}^T$. Then, the Fr\'echet mean of the Cholesky decomposed matrices $\mathcal{L}$ is given by  

\begin{align}
\mathcal{F}_{\textsc{cholesky}} &= \frac{1}{n} \sum_{i = 1}^{n} \lfloor\mathcal{L}_i\rfloor \hspace{0.3cm} +\nonumber \\
&\quad \exp\left(\frac{1}{n} \sum_{i = 1}^{n} \log(\mathbb{D}(\mathcal{L}_i)\right). \label{eq:mean}
\end{align}

\subsubsection{{\em k}-medoids clustering algorithm}
\label{sec:kMedoids}

 We implement the classic {\em k}-medoids algorithm \cite{Kaufman1990PAM} using partitioning around medoids (PAM) heuristic by replacing the Euclidean distance with the geodesic distance in equation \ref{eq:distance}.

\subsubsection{Minimum distance to mean algorithm (MDM)}
\label{sec:MDM}
Given {\em m} classification classes and {\em n} training samples, SPD matrices in the training set $\{\mathcal{E}_i^j\}$, where $i\in\{1, 2, ..., $ {\em n}\}and $j\in\{1, 2, ..., $ {\em m}\} are used to construct centroids for each of the {\em m} classes such that the centroid of class $j$ is,
\begin{equation}
    \mathcal{C}^j = \mathbb{E}(\{\textsc{cholesky}({\mathcal{E}}^j)\}), 
\end{equation}
\noindent where the Fr\'echet mean $\mathbb{E}$ is calculated according to equation \ref{eq:mean}. Given a test dataset of SPD matrices $\{\mathcal{T}\}$, $T\in\mathcal{T}$ is assigned to that class whose centroid is nearest to $\textsc{cholesky}(T)$. That is, the class of $T$ is
\begin{equation}
    \arg\min_j d(\textsc{cholesky}(T), \mathcal{C}^j),
\end{equation}
where $d(.)$ is calculated according to equation \ref{eq:distance}.

\subsubsection{Support Vector Machine (SVM)}
\label{sec:SVM}
For training the SVM, we use a kernel
\[
\mathscr{K}: \textsc{Cholesky space} \times \textsc{cholesky space}\rightarrow \mathbb{R}, 
\]
such that
\begin{equation}
	\mathscr{K} = \exp(-\gamma d^2(L_1, L_2)),
 \label{eq: SVM}
\end{equation}
where $L_1$, $L_2$ are Cholesky factors of SPD matrice $\mathcal{E}_1$ and $\mathcal{E}_2$ and $\gamma > 0$. $d(.)$ is defined according to equation \ref{eq:distance}. In appendix \ref{apd:A}, following the arguments in \citeauthor{7063231}, we prove that the kernel in equation \ref{eq: SVM} is a valid kernel.

\subsubsection{{\em t-}SNE for EMG data visualization}
\label{sec:tSNE}
We use the {\em t}-SNE ({\em t}-Stochastic Neighbor Embedding) algorithm, adapted from \citeauthor{van2008visualizing}, for data visualization. Unlike standard {\em t}-SNE, which takes vectors as input and uses Euclidean distance, we input edge matrices $\mathcal{E}$ and employ the distance defined in equation \ref{eq:distance}.
\section{Results}
Here, we demonstrate that different hand gestures can be decoded with high accuracy using simple algorithms such as {\em k}-medoids ({\em section}~\ref{sec:kMedoids}), minimum distance to mean ({\em section}~\ref{sec:MDM}), and SVM ({\em section}~\ref{sec:SVM}) on the manifold of SPD matrices. This finding highlights that EMG signal covariance matrices capture meaningful and distinguishing information, and that the manifold of SPD matrices serves as a natural embedding space for EMG signals.

\subsection{Results for \textsc{UCD-MyoVerse-Hand-0}}
EMG data was collected at a sampling frequency of 2000 Hz using 12 electrodes, with each gesture lasting 2 {\em s}. For each gesture, the EMG data is represented as a matrix $\mathcal{X}$ of dimensions $(12 \times 4000)$, where 12 corresponds to the number of electrode channels and 4000 corresponds to the temporal dimension (i.e., 2 {\em s} $\times$ 2000 Hz). The corresponding covariance SPD matrices, denoted as $\mathcal{E}$, have dimensions $(12 \times 12)$. We decode covariance matrices $\mathcal{E}$ using MDM, SVM, and {\em k-}medoids algorithms defined on the manifold of SPD matrices.

We present the average classification accuracies across subjects in table~\ref{tab:UCDMyoVerseR}. Detailed subject-wise results are provided in appendix~\ref{apd:res1}.

\begin{table}[htbp]
\centering
\footnotesize
\renewcommand{\arraystretch}{0.9}
\begin{tabular}{|c|c|c|c|}
    \hline
    \multirow{2}{*}{\textbf{\scriptsize Subject Number}} & \multicolumn{3}{c|}{\textbf{\scriptsize Classification Methods}} \\
    \cline{2-4}
    & \textbf{\scriptsize MDM} & \textbf{\scriptsize SVM ($\gamma = 1$)} & \textbf{\scriptsize {\em k}-medoids} \\
    \hline
    \textbf{\scriptsize Average} & {\scriptsize 0.82} & {\scriptsize 0.86} & {\scriptsize 0.70} \\
    \hline
\end{tabular}
\caption{Average classification accuracy for 30 subjects in \textsc{UCD-MyoVerse-Hand-0}. Data from the first three sessions (i.e., the first 18 repetitions of each gesture) were used for training, while data from the last three sessions (i.e., the last 18 repetitions of each gesture) were used for testing the MDM and SVM algorithms. Chance decoding accuracy is $\frac{1}{10} = 0.1$.}
\label{tab:UCDMyoVerseR}
\end{table}

\subsection{Results for \textsc{Ninapro: Database 2-Exercise 1}}
EMG data was collected at a sampling frequency of 2000 Hz using 12 electrodes, with each gesture lasting 5 {\em s}. For each gesture, the EMG data is represented as a matrix $\mathcal{X}$ with 12 rows (corresponding to the 12 electrode channels) and 10000 columns (corresponding to the temporal dimension, i.e., 5 {\em s} $\times$ 2000 Hz). Therefore, the dimensions of $\mathcal{X}$ are $(12 \times 10000)$.

The corresponding covariance SPD matrices, denoted as $\mathcal{E}$, have dimensions $(12 \times 12)$. We decode these covariance matrices using the MDM, SVM, and {\em k}-medoids algorithms, which are defined on the manifold of SPD matrices.

To evaluate our approach, we compare the results obtained using our methods with those reported in \citeauthor{9463745} and \citeauthor{rahimian2021temgnet} (table~\ref{tab:ninaproComparison}). Detailed subject-wise results for all 40 subjects using manifold-based methods are provided in appendix~\ref{apd:res2}. For {\em train-test} data split, refer to \citeauthor{9463745, rahimian2021temgnet}

\begin{table}[H]
\renewcommand{\arraystretch}{0.83}
\centering
\begin{tabular}{|c|c|c|}
\hline
\multicolumn{1}{|c|}{} & {\footnotesize Method} & {\footnotesize Accuracy} \\
\hline
& {\scriptsize 4-layer 3rd order dilation} & {\scriptsize 0.824} \\
& {\scriptsize 4-layer 3rd order dilation (LSTM)} & {\scriptsize 0.797} \\
& {\scriptsize 4-layer 2nd order dilation (LSTM)} & {\scriptsize 0.796} \\
& {\scriptsize 4-layer 1st order dilation (LSTM)} & {\scriptsize 0.793} \\
{\scriptsize \citeauthor{9463745}} & {\scriptsize 4-layer baseline} & {\scriptsize 0.753} \\
& {\scriptsize 2-layer CNN}  & {\scriptsize 0.746} \\
& {\scriptsize 2-layer LSTM} & {\scriptsize 0.702} \\
& {\scriptsize 1-layer LSTM} & {\scriptsize 0.684} \\
& {\scriptsize 2-layer MLP} & {\scriptsize 0.662} \\
& {\scriptsize SVM (Euclidean)} & {\scriptsize 0.307} \\
\hline
& {\scriptsize TEMGNet {\em 200} m{\em s}} & {\scriptsize 0.821} \\
{\scriptsize \citeauthor{rahimian2021temgnet}} & {\scriptsize TEMGNet {\em 300} m{\em s}} & {\scriptsize 0.829} \\
\hline
& {\scriptsize Manifold MDM} & {\scriptsize 0.92} \\
{\scriptsize Proposed} & {\scriptsize Manifold SVM} & {\scriptsize 0.93} \\
{\scriptsize methods} & {\scriptsize Manifold {\em k}-medoids} & {\scriptsize 0.82} \\
\hline
\end{tabular}
\caption{Our proposed methods leverage manifold representation 
and outperform baselines. The unsupervised {\em k}-medoids 
algorithm is, to our knowledge, the only unsupervised 
method for EMG classification. Unlike neural networks 
with tens of thousands of parameters, our methods 
are computationally efficient. Chance decoding accuracy is $\frac{1}{17} = 0.059$.}
\label{tab:ninaproComparison}
\end{table}

\subsection{Results for high density EMG in \textsc{\citeauthor{Malešević2021}}}
High-density EMG was collected at 2048 Hz using 128 electrodes. The duration of each gesture varied, which we denote as the variable $\mathcal{T}$. The data corpus consisted of 65 unique hand gestures, with each gesture repeated five times. For each gesture, the EMG data is represented as a matrix $\mathcal{X}$, where $\mathcal{X}$ has $\mathcal{T}$ columns (temporal dimension) and 128 rows (corresponding to the 128 electrode channels). Thus, the dimensions of $\mathcal{X}$ are $(128 \times \mathcal{T})$.

The covariance SPD matrix $\mathcal{E}$ has dimensions $(128 \times 128)$. We use MDM, SVM, and {\em k}-medoids algorithms, as described previously, to classify the SPD matrices $\mathcal{E}$. The results of our manifold-based classification methods are presented in table \ref{tab:highDensity}.

Additionally, we compare our results with the methods proposed by \citeauthor{Montazerin2023} in table \ref{tab:highdensityComparison}.

\begin{table}[htbp]
\footnotesize
\renewcommand{\arraystretch}{0.95}
	\centering
	\label{tab:table4}
	\begin{tabular}{|c|c|c|c|}
		\hline
		\multirow{2}{*}{\scriptsize Fold number} & \multicolumn{3}{c|}{\scriptsize Classification methods} \\
		\cline{2-4}
		& {\scriptsize MDM} & {\scriptsize SVM ($\gamma = 0.1$)} &  {\scriptsize {\em k}-medoids}\\
		\cline{1-4}
		{\scriptsize 1} & {\scriptsize 0.83 $\pm$ 0.07} & {\scriptsize 0.84 $\pm$ 0.08} & \\
		\cline{1-3}
		{\scriptsize 2} & {\scriptsize 0.95 $\pm$ 0.04} & {\scriptsize 0.96 $\pm$ 0.03} & \\
		\cline{1-3}
		{\scriptsize 3} & {\scriptsize 0.96 $\pm$ 0.03} & {\scriptsize 0.97 $\pm$ 0.03} & {\scriptsize 0.87 $\pm$ 0.05}\\
		\cline{1-3}
		{\scriptsize 4} & {\scriptsize 0.95 $\pm$ 0.04} & {\scriptsize 0.96 $\pm$ 0.04} & \\
		\cline{1-3}
		{\scriptsize 5} & {\scriptsize 0.91 $\pm$ 0.05} & {\scriptsize 0.92 $\pm$ 0.04} & \\
		\hline
		
		\hline
		{\bf \scriptsize Average} &  \bf{\scriptsize 0.92 $\pm$ 0.07} & {\scriptsize \bf0.93 $\pm$ 0.07} & \bf{\scriptsize 0.87 $\pm$ 0.05}\\
		\hline
	\end{tabular}
	
	\caption{Classification accuracy averaged across {\em 19} subjects in \citeauthor{Malešević2021}  Following the work in \citeauthor{Montazerin2023}, we perform 5-fold cross validation analysis. Chance decoding accuracy is $\frac{1}{65}$.}\label{tab:highDensity}
\end{table}
\begin{table}[htbp]
\footnotesize
\renewcommand{\arraystretch}{0.8}
\centering
\begin{tabular}{|c|c|c|}
\hline
\multicolumn{1}{|c|}{} & {\scriptsize Method} & {\scriptsize Accuracy} \\
\hline
{\scriptsize \citeauthor{Montazerin2023}} & {\scriptsize {\em 512} sample window and}  & {\scriptsize 0.92} \\
& {\scriptsize {\em 128} channels ({\em their best})} & \\
\hline
& {\scriptsize Manifold MDM} & {\scriptsize 0.92} \\
{\scriptsize Proposed} & {\scriptsize Manifold SVM} & {\scriptsize 0.93} \\
{\scriptsize methods} & {\scriptsize Manifold {\em k}-medoids} & {\scriptsize 0.87} \\
\hline
\end{tabular}
\caption{The proposed methods achieve performance comparable to the transformer-based approach in \citeauthor{Montazerin2023}, while being computationally efficient. Moreover, they are better suited for deployment across individuals and real-time adaptation due to their transparent nature. For instance, drift in EMG signals caused by sensor position shifts or muscle fatigue can be addressed simply by updating the centroids of the SPD matrices in the manifold MDM algorithm, rather than repeatedly fine-tuning or retraining a transformer. Chance decoding accuracy is $\frac{1}{65}$.}
\label{tab:highdensityComparison}
\end{table}
\subsection{EMG data visualization using {\em t-}SNE}
We use {\em t}-SNE ({\em section} \ref{sec:tSNE}) to visualize the SPD covariance matrices of hand gestures. These visualizations demonstrate that the covariance matrices of EMG signals exhibit structured representations on the manifold of SPD matrices, a finding further validated through simple classification methods defined on the manifold in previous sections.

\begin{figure}[h!]
	\centering
	\includegraphics[width=0.8\textwidth]{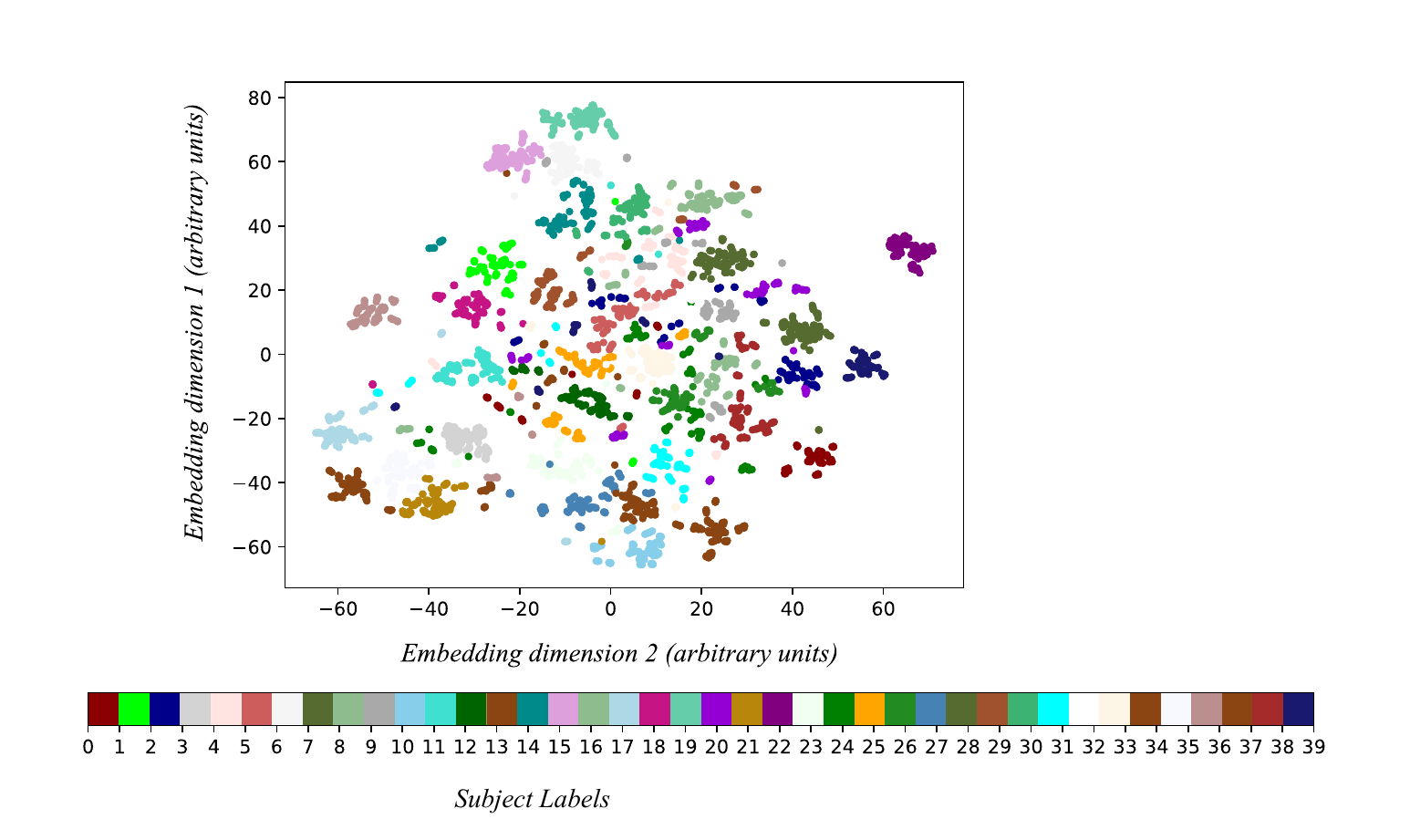}
	\caption{{\em t}-SNE of SPD covariance matrices using Riemannian distance indicates that the SPD matrices from different subjects lie in different neighborhoods of the manifold. This is due to shift in EMG signals owing to the combined effect of various anatomical, physiological, and circumstantial factors. Embedding is for \textsc{Ninapro: Database 2-Exercise 1}. Embedding is colored according to subjects.  Each of the 40 subjects performed 102 trials (17 gestures, each repeated 6 times).}
	\label{fig:tSNE40Subjects}
\end{figure}

\begin{figure}[h!]
	\centering
	\includegraphics[width=.6\textwidth]{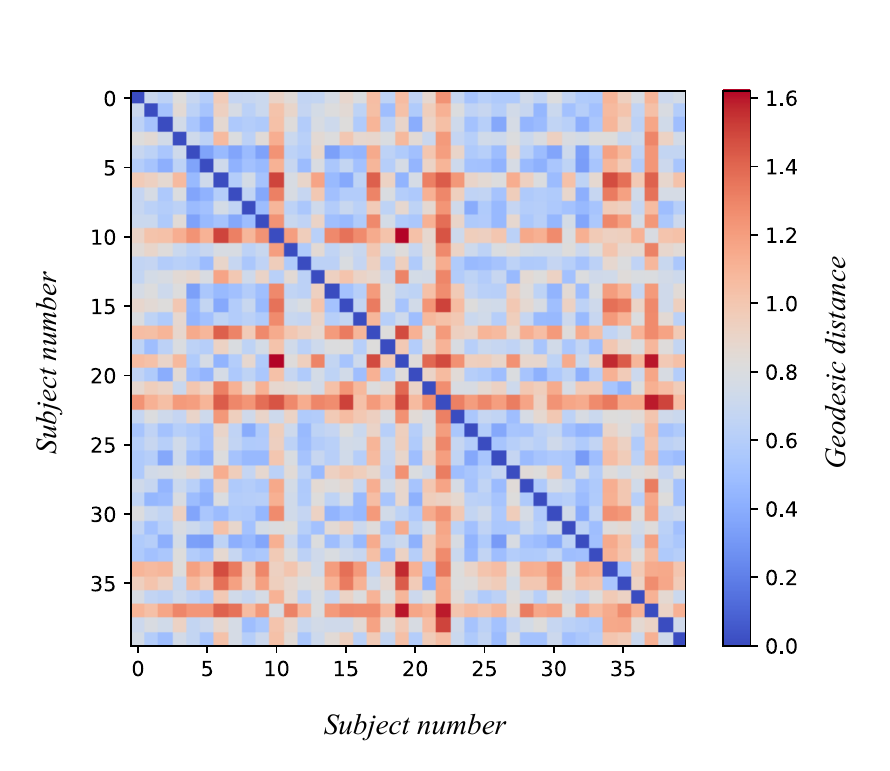}
	\caption{\footnotesize Riemannian geodesic distance between the centroids of SPD covariance matrices of 40 subjects in \textsc{Ninapro: Database 2-Exercise 1}. Geodesic distance quantifies the differences in EMG signals between subjects due to the combined effect of various physiological and anatomical factors. The centroid of a given subject is calculated as the {\em Log-Cholesky} average of SPD covariance matrices of all 102 trials (17 gestures, each repeated 6 times). {\em X} and {\em Y} axes are numbered according to subjects.}
	\label{fig:subjectDistances}
\end{figure}

The {\em t}-SNE embedding reveals that SPD matrices from different subjects occupy distinct neighborhoods on the manifold, as shown in figure \ref{fig:tSNE40Subjects}. This variation in EMG signals arises from a combination of anatomical, physiological, and circumstantial factors and can be quantified using a single geodesic distance metric across all individuals, as depicted in figure \ref{fig:subjectDistances}. Due to these differences in spatial muscle contraction patterns among individuals, generalizing deep learning algorithms across the entire population remains challenging.

\begin{figure}[h!]
	\centering
	\includegraphics[width=.8\textwidth]{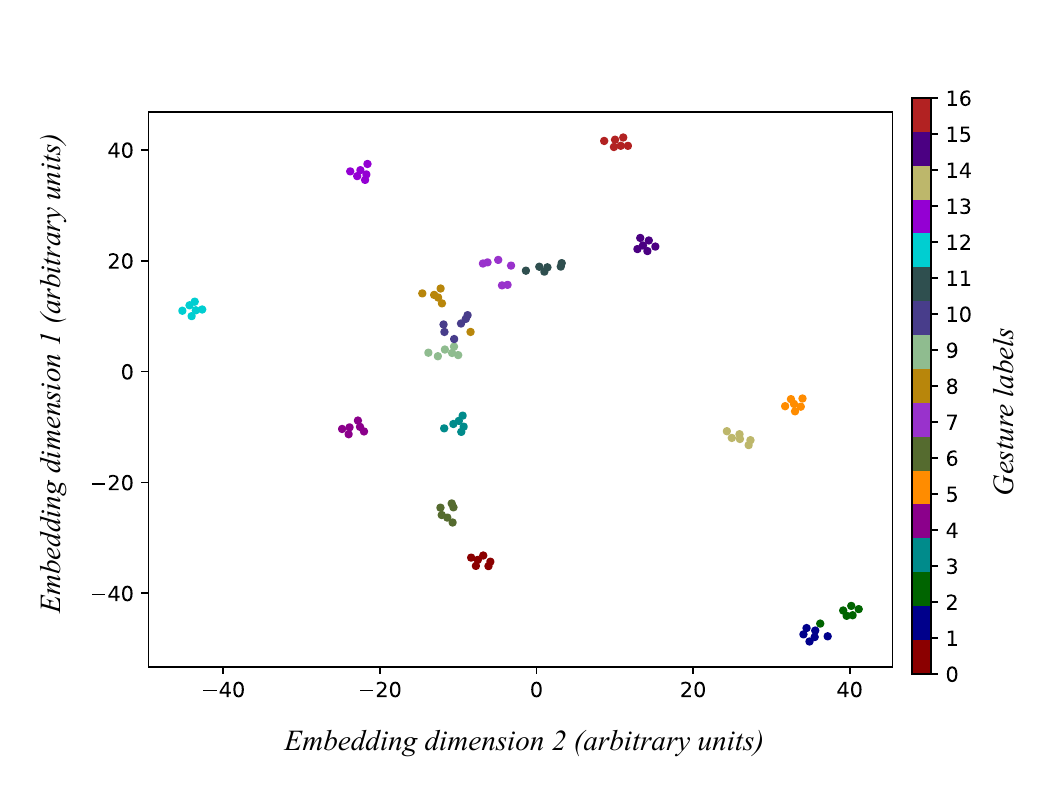}
	\caption{\footnotesize {\em t}-SNE of SPD matrices of {\em subject 0} from  \textsc{Ninapro: Database 2-Exercise 1} shows that different gestures within a subject have contrasting spatial patterns. We can classify these distinct gestures with supervised algorithms such as minimum distance to mean (MDM) and support vector machine (SVM) or unsupervised algorithms such as {\em k}-medoids clustering using Riemannian distance. Classification accuracy using the above methods is presented in appendix \ref{apd:res2} for all 40 subjects. Embedding is colored according to gestures. The subject performed 17 gestures with each gesture repeated six times.\\
    \noindent The gestures are: 0: thumb up, 1: extension of index and middle - flexion of the others,  2: flexion of ring and little finger - extension of the others, 3: thumb opposing base of little finger, 4: abduction of all fingers, 5: fingers flexed together in fist, 6: pointing index, 7: adduction of extended fingers, 8: wrist supination (axis: middle finger), 9: wrist pronation (axis: middle finger), 10: wrist supination (axis: little finger), 11: wrist pronation (axis: little finger), 12: wrist flexion, 13: wrist extension, 14: wrist radial deviation, 15: wrist ulnar deviation, and 16: wrist extension with closed hand.}
  \label{fig:subject0Ninapro}
\end{figure}

However, with this appropriate non-Euclidean representation of the data, spatial muscle contraction patterns within an individual are sufficiently distinct to enable unsupervised classification of various gestures, as demonstrated in figure \ref{fig:subject0Ninapro}, thereby eliminating the need for complex transfer learning paradigms or model retraining when deploying across individuals. 

Furthermore, since patterns from different individuals occupy separate neighborhoods on the manifold, directly comparing gestures across all subjects leads to extremely poor gesture differentiation (figure \ref{fig:40SubjectColorByGestures}).

Nevertheless, using parallel transport (see appendix \ref{apd:B}), these distinct gesture patterns can be aligned across individuals, with an accuracy bounded by the accuracy of unsupervised classification (figure \ref{fig:align}). Figure \ref{fig:Cone} illustrates why {\em parallel} transport (see appendix \ref{apd:B}), as opposed to Euclidean transport, is the appropriate approach in the manifold space.

\section{Observations and conclusions}
\textcircled{\scriptsize 1} EMG signals are multivariate and capture muscle activation patterns across spatially separated locations using multiple sensor nodes. This multivariate nature of EMG, which reflects a distributed yet concerted effort from multiple muscles to execute a given gesture, suggests that the data structure is not inherently Euclidean. In this article, we empirically demonstrate that this is indeed the case.

Signal covariance matrices, which represent the covariances of EMG signals across all pairs of sensor nodes, form structured representations on the manifold of SPD matrices, naturally distinguishing different hand gestures. This suggests that covariance features, rather than convolutional filters, are more suitable for EMG signal feature extraction.

\textcircled{\scriptsize 2} We showed that EMG signals exhibit discriminative representations on the manifold, allowing classification even with unsupervised clustering algorithms. However, EMG signals from different individuals occupy distinct neighborhoods within the manifold space.

We need to verify whether EMG signals from certain individuals or groups - such as those with a high BMI, who typically have more subcutaneous fat - lie in the same neighborhood or whether each individual is inherently different. If the latter is true, we must further explore its implications for generalizable machine learning models.

\textcircled{\scriptsize 3} Following the previous point, we should explore whether zero-shot learning algorithms can be developed, allowing a model trained on a set of individuals to generalize to unseen individuals.

As demonstrated here, few-shot learning is straightforward. If we have a few labeled samples for each gesture, we can perform unsupervised classification and use parallel transport to align EMG signals across individuals.

\onecolumn

\begin{figure}[h!]
	\centering
	\includegraphics[width=.8\textwidth]{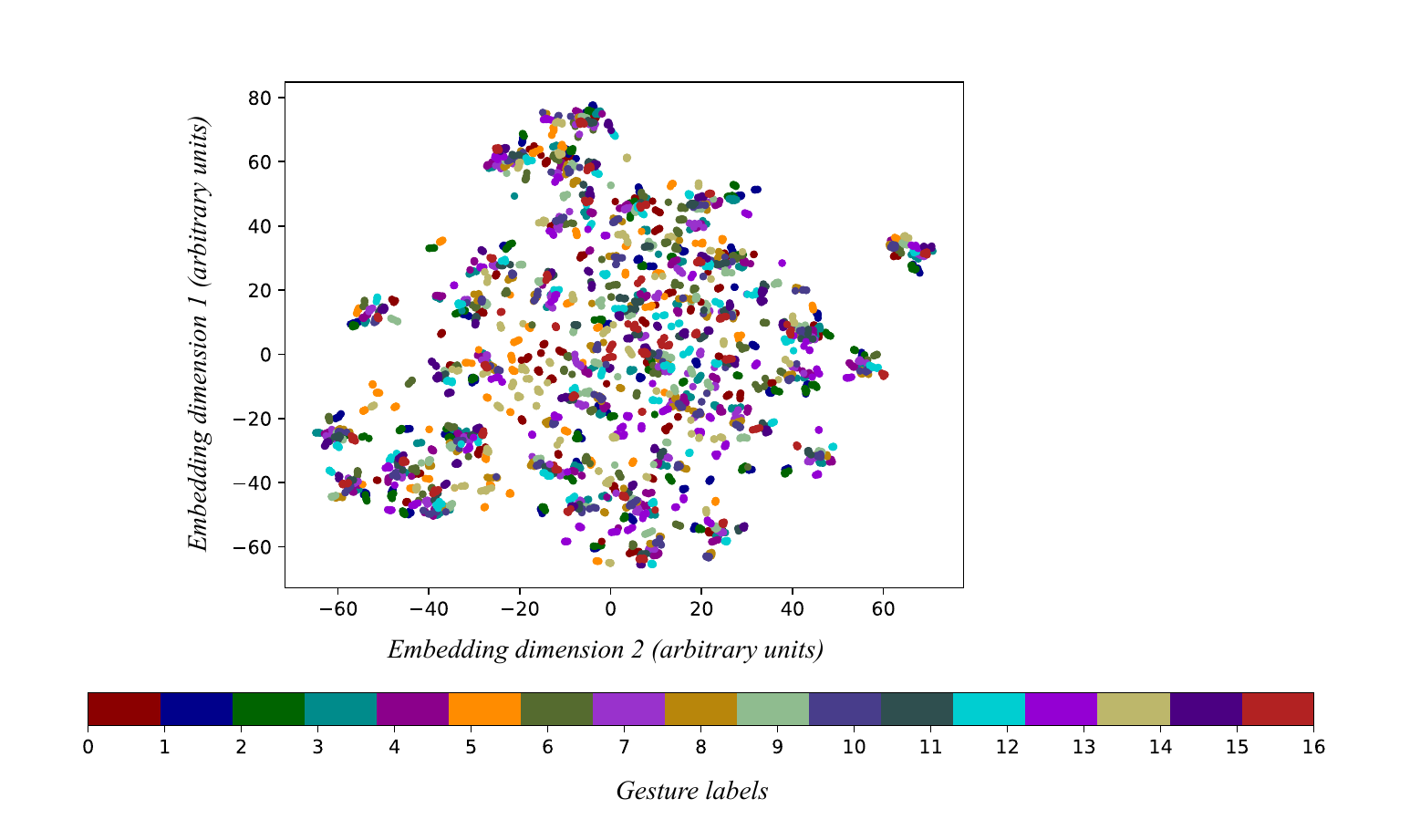}
	\caption{\footnotesize{\em t}-SNE of SPD covariance matrices using Riemannian distance without accounting for intersubject differences reveals that the SPD matrices for the same gesture from different subjects do not cluster together. This is the same embedding in figure \ref{fig:tSNE40Subjects} colored according to gestures (instead of subjects).  Each of the 40 subjects performed 102 trials (17 gestures, each repeated 6 times). \\\noindent  The gestures are: 0: thumb up, 1: extension of index and middle - flexion of the others,  2: flexion of ring and little finger - extension of the others, 3: thumb opposing base of little finger, 4: abduction of all fingers, 5: fingers flexed together in fist, 6: pointing index, 7: adduction of extended fingers, 8: wrist supination (axis: middle finger), 9: wrist pronation (axis: middle finger), 10: wrist supination (axis: little finger), 11: wrist pronation (axis: little finger), 12: wrist flexion, 13: wrist extension, 14: wrist radial deviation, 15: wrist ulnar deviation, and 16: wrist extension with closed hand.}
	\label{fig:40SubjectColorByGestures}
\end{figure}

\begin{figure}[h!]
	\centering
	\includegraphics[width=.8\textwidth]{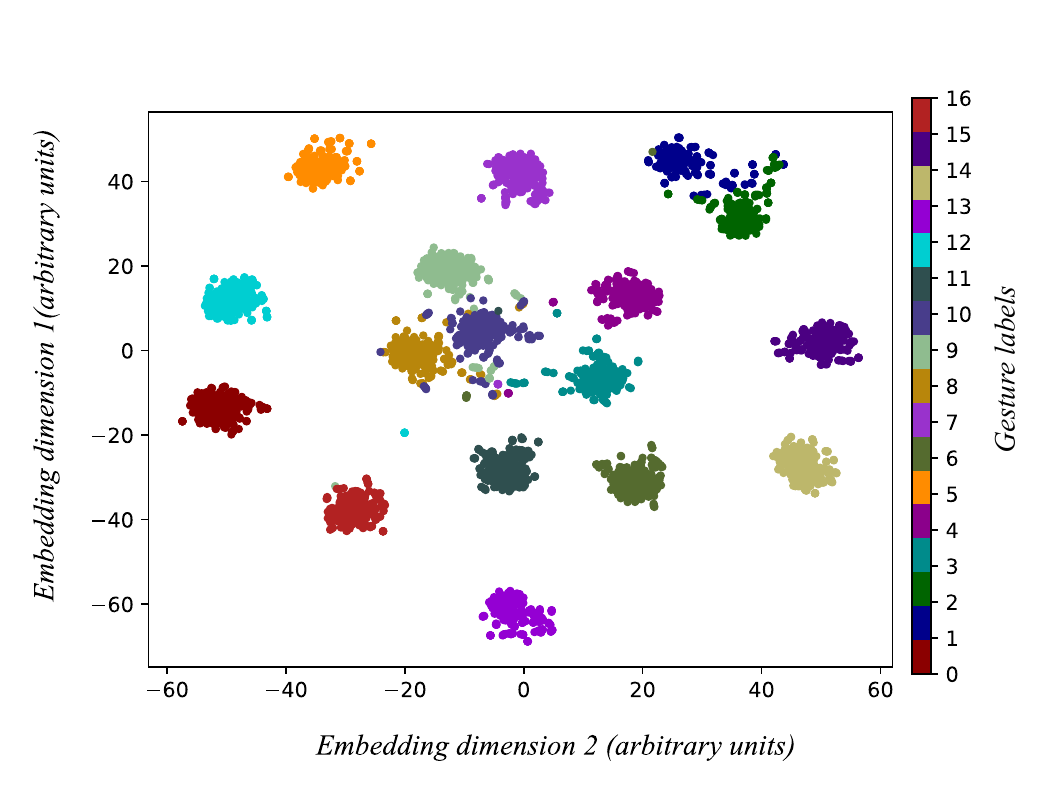}
	\caption{\footnotesize {\em t}-SNE of SPD covariance matrices using Riemannian distance after aligning the gestures using {\em parallel transport} as described in appendix \ref{apd:B} and illustrated in figure \ref{fig:Cone}. Data is aligned assuming ground truth labels for visualization purposes. Otherwise, alignment accuracy is bounded by the accuracy of unsupervised classification. This demonstrates how data from different individuals in figure \ref{fig:40SubjectColorByGestures} can be aligned according to gestures so that the same gestures from different individuals are transported to the same neighborhood in the manifold space.
		Embedding is colored according to gestures.  Each of the 40 subjects performed 102 trials (17 gestures, each repeated 6 times). 
		\\\noindent The gestures are: 0: thumb up, 1: extension of index and middle - flexion of the others,  2: flexion of ring and little finger - extension of the others, 3: thumb opposing base of little finger, 4: abduction of all fingers, 5: fingers flexed together in fist, 6: pointing index, 7: adduction of extended fingers, 8: wrist supination (axis: middle finger), 9: wrist pronation (axis: middle finger), 10: wrist supination (axis: little finger), 11: wrist pronation (axis: little finger), 12: wrist flexion, 13: wrist extension, 14: wrist radial deviation, 15: wrist ulnar deviation, and 16: wrist extension with closed hand.}
	\label{fig:align}
\end{figure}

\begin{figure}[h!]
	\centering
	\includegraphics[width=.4\textwidth]{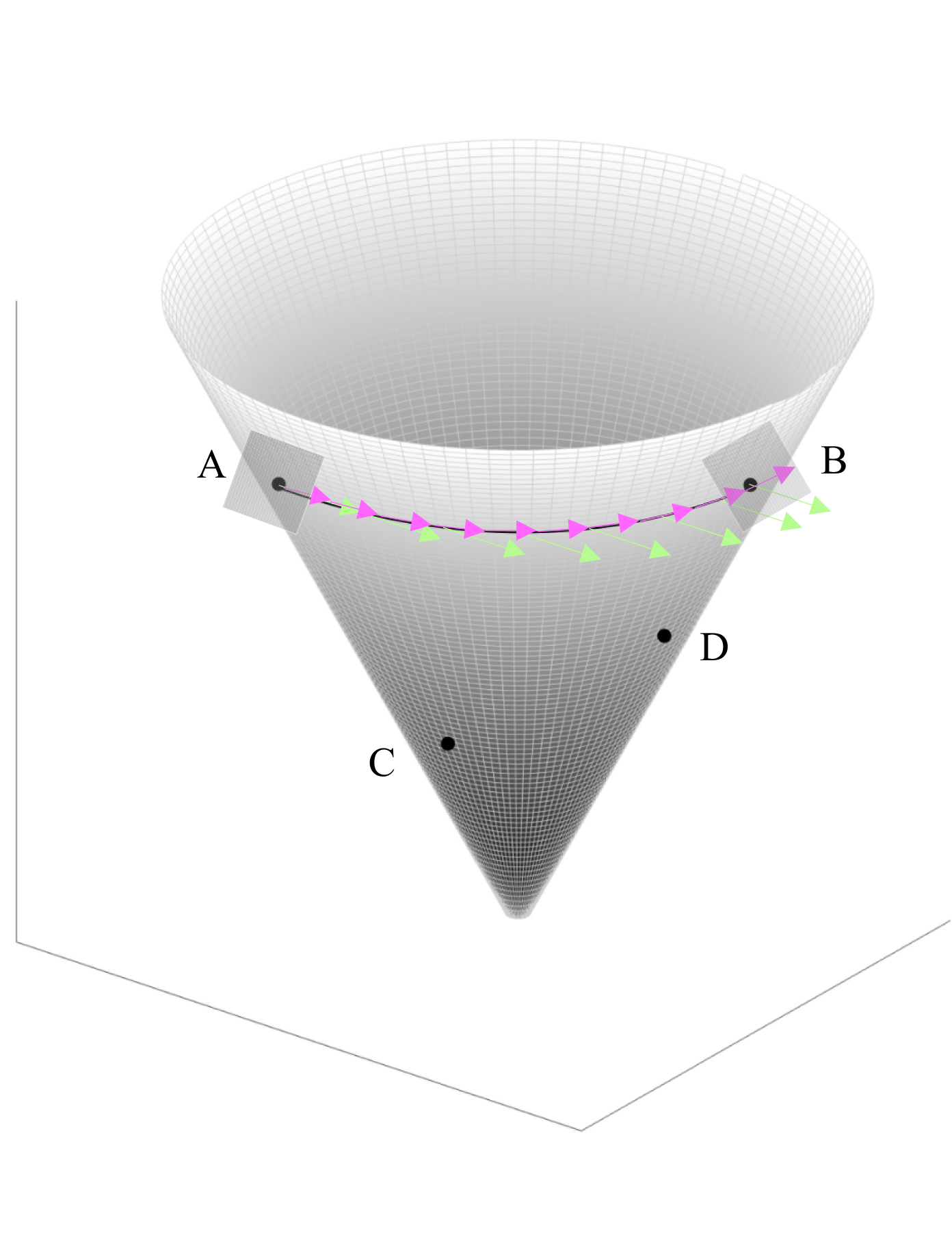}
	\caption{ \footnotesize Euclidean transport is inappropriate in the manifold space.  Point $A$ is an SPD matrix  that represents the EMG signal covariance for a given gesture, say a pinch, for a given subject. Point $B$ is another SPD matrix representing the reference location for that gesture on the manifold: say, a particular reference subject's centroid for the pinch gesture. The length of the curve joining $A$ and $B$ is the geodesic distance between the two points. Points $C$ and $D$, like $A$ and $B$ are SPD matrices that represent different gestures; for instance, the reference locations for a power grasp and a wrist flexion. {\em Green} arrows illustrate the Euclidean transport of a tangent vector from $A$ to $B$. {\em Pink} arrows represent the {\em parallel} transport of a tangent vector from $A$ to $B$. Parallel transport rotates the vector along the path from $A$ to $B$ so that the vector in the tangent plane of $A$ remains in the tangent plane of $B$ when transported. It can be seen that the Euclidean transport of a vector in the tangent plane of $A$ results in vectors that are no longer in the tangent plane along the path to $B$. The Euclidean transport of vectors therefore cannot appropriately traverse a geodesic. The {\em square patches} at $A$ and $B$ represent the tangent planes.}	\label{fig:Cone}
\end{figure}
\begin{figure}[h!]
	\centering
	\includegraphics[width=.8\textwidth]{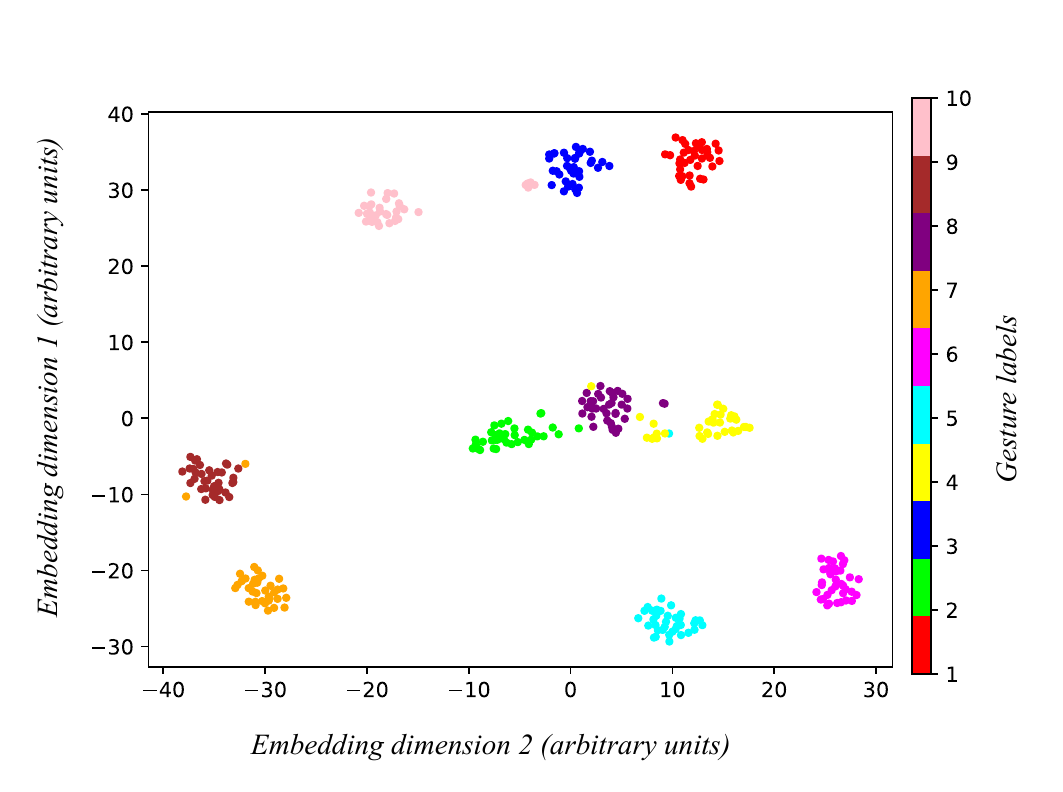}
	\caption{\footnotesize {\em t}-SNE of SPD matrices of {\em subject 0} from \textsc{UCD-MyoVerse-Hand-0} shows that different gestures within a subject have contrasting spatial patterns. We can classify these distinct gestures with supervised algorithms such as minimum distance to mean (MDM) and support vector machine (SVM) or unsupervised algorithms such as {\em k}-medoids clustering using Riemannian distance. Average classification accuracy across subjects using the above methods is presented in table \ref{tab:UCDMyoVerseR}. Embedding is colored according to gestures. The subject performed 10 gestures with each gesture repeated 36 times.\\
\noindent Ten gestures are: 1 - Down, 2 - Index finger pinch, 3 - Left, 4 - Middle finger pinch, 5 - Index point, 6 - Power grasp, 7 - Right, 8 - Two finger pinch, 9 - Up, 10 - Splay.}
	\label{fig:UCDMyoVerse}
\end{figure}
\twocolumn
\paragraph*{Ethical statement}
Research was conducted in accordance with the principles embodied in the Declaration of Helsinki and in accordance with the University of California, Davis Institutional Review Board Administration protocol 2078695-1. All participants provided written informed consent. Consent was also given for publication of the deidentified data by all participants. Participants were healthy volunteers and were selected from any gender and all ethnic and racial groups. Subjects were aged 18 or above, were able to fully understand spoken and written English, and were capable of following task instructions. Subjects had no skin conditions or wounds where electrodes were placed. Subjects were excluded if they had uncorrected vision problems or neuromotor disorders that prevented them from articulating speech. Children, adults who were unable to consent, and prisoners were not included in the experiments. 
\paragraph*{Acknowledgements}
This work was supported by Meta Platforms Technologies (Facebook Research) with an award to L.M.M through the Ethical
Neurotechnology program and by the UC Davis School of Medicine Cultivating Team Science Award to L.M.M. We would like to thank Stephanie Naufel at Facebook Reality Labs for her valuable guidance. We thank Carlos Carrasco, Marcus Battraw, and Jonathon Schofield for their advice in designing the experiment and collection of data. We also thank Neha Kaul, Safa Amar, and Saniya Kotwal for their help in collection of data.
\paragraph*{Author Contributions}
L. M. M. conceptualized the study and framed the problem statements. H. T. G. formulated the mathematical framework, implemented the codes, and collected the data. H. T. G. and L. M. M. authored and reviewed the manuscript. 
\section*{Competing Interests}
The authors declare no competing interests.


\appendix
\section{Manifold SVM kernel}
\label{apd:A}
Here, we prove that the kernel used in {\em section} \ref{sec:SVM} is a valid kernel.

We shall prove that the kernel $\mathscr{K} = \exp(-\gamma d^2(L_1, L_2))$ for $L_1$, $L_2\in\textsc{Cholesky space}$ is a valid kernel for all $\gamma > 0$. 

For a kernel to be valid, it must be positive definite \cite{7063231}. 

From {\em theorem 5.2} in \citeauthor{7063231}, a kernel 
\begin{multline*}
    \mathscr{K}: \textsc{Cholesky space} \times \textsc{cholesky space}\rightarrow \mathbb{R} \\:= \exp(-\gamma d^2(L_1, L_2)) 
\end{multline*}
is positive definite for all $\gamma > 0$ if and only if $d^2(L_1, L_2)$ is negative definite. 

From {\em lemma 5.5} in \citeauthor{7063231}, if 
\[
\psi:\textsc{Choleasky space}\rightarrow F, 
\]
where $F$ is the Frobenius inner product space is a function, then, 
\begin{multline*}
    \mathscr{O}:\textsc{Cholesky space}\times\textsc{Cholesky space}\rightarrow\mathbb{R} 
\end{multline*}
defined by 
\[
    \mathscr{O}(L_1, L_2) = ||\psi(L_1) - \psi(L_2)||^2_F
\]
is negative definite. 

Following {\em theorem 6.1} \cite{7063231},  
\begin{multline*}
    \mathscr{K}:\textsc{Cholesky space} \times \textsc{cholesky space}\rightarrow \mathbb{R}
\end{multline*}
given by 
\[
\mathscr{K}(L_1, L_2) = \exp(-\gamma d^2(L_1, L_2))
\]
is a positive definite kernel for all $\gamma > 0$ if and only if there exists a function 
\[
\psi:\textsc{Cholesky space}\rightarrow F
\]
such that 
\[
d(L_1, L_2) = ||\psi(L_1) - \psi(L_2)||_F. 
\]
We can now define 
\[
\psi = \lfloor L\rfloor + \log \mathbb{D}(L)
\]
for all  $L\in$ \textsc{Cholesky space}.

We have 
\begin{multline*}
    d(L_1, L_2) = \{||\lfloor L_1\rfloor - \lfloor L_2\rfloor||_F^2 + \\||\log\mathbb{D}(L_1) - \log\mathbb{D}(L_2)||_F^2\}^{1/2}.
\end{multline*}
\begin{multline*}
    ||\psi(L_1) - \psi(L_2)||_F = \\||\lfloor L_1 \rfloor + \log\mathbb{D}(L_1) - \lfloor L_2 \rfloor - \log\mathbb{D}(L_2)||_F \\= \{||\lfloor L_1 \rfloor - \lfloor L_2\rfloor||_F^2+||\log\mathbb{D}(L_1) - \log\mathbb{D}(L_2)||_F^2\}^{1/2}
\end{multline*}
Therefore, 
\[
d(L_1, L_2) = ||\psi(L_1) - \psi(L_2)||_F.
\]
Hence, the kernel 
\[
\mathscr{K} = \exp{(-\gamma d^2(L_1, L_2))}
\]
is a valid kernel.

In the above, $d(.)$ is defined according to equation \ref{eq:distance} and $||.||_F$ is the Frobenius norm.

\section{Parallel transport}
\label{apd:B}
Here, we explain parallel transport in detail.

As given in \citeauthor{Lin_2019}, the logarithm map from the manifold to its tangent space at $L$ is given by
\begin{equation}
	\widetilde{\text{Log}}_LK = \lfloor K\rfloor - \lfloor L\rfloor + \mathbb{D}(L)\log\{\mathbb{D}(L)^{-1}\mathbb{D}(K)\},
    \label{eq:tanmap}
\end{equation}
where $L$, $K\in \textsc{Cholesky space}$ and $\widetilde{\text{Log}_L}K\in \textsc{Tangent space}$

The exponential map from the tangent space to the manifold space is given by,
\begin{equation}
	\widetilde{\text{Exp}}_LX = \lfloor L\rfloor + \lfloor X\rfloor + \mathbb{D}(L)\exp\{\mathbb{D}(X)\mathbb{D}(L)^{-1}\},
    \label{eq:expmap}
\end{equation}
where $L$, $\widetilde{\text{Exp}}_LX \in \textsc{Cholesky space}$ and $X\in \textsc{Tangent space}$.

A tangent vector $X\in \textsc{Tangent space}$ is parallelly transported to the tangent vector 
\begin{equation}
	\lfloor X \rfloor + \mathbb{D}(K)\mathbb{D}(L)^{-1}\mathbb{D}(X)
    \label{eq:pt}
\end{equation}
at $K$ \cite{Lin_2019}.\\\vspace{0.2cm}

\noindent Given a set of subjects ({\em S}) and a set of gestures ({\em G}), let us denote gestures of a particular hand movement (Cholesky factorized covariance SPD matrices) belonging to a particular subject as $\mathcal{G}_s^g$, {\em s} $\in$ {\em S} and {\em g} $\in$ {\em G}. Parallel transport algorithm is as defined below. 

\begin{algorithm}
	\caption{Parallel transport}\label{alg:cap}
	\begin{algorithmic}
		\STATE \textbf{Input:} Cholesky matrices $\mathcal{G}_s^g$
		\STATE \textbf{Output:} Parallelly transported Cholesky matrices.
        \vspace{0.25cm}
        
		\STATE Compute the Riemannian mean \\
        $\bar{\mathcal{G}}_s^ g = \mathbb{E}(\text{all elements in } \mathcal{G}_s^g)$ using equation \ref{eq:mean}.
        \vspace{0.25cm}
        
		\STATE Map all elements in ${\mathcal{G}_s^g}$ to the tangent space at $\bar{\mathcal{G}_s^g}$ using equation \ref{eq:tanmap}. \\
        Denote this mapped set as $\textsc{tangent mapped}(\mathcal{G}_s^g)$.
        \vspace{0.25cm}
        
		\STATE Parallelly transport the tangent vectors in $\textsc{tangent mapped}(\mathcal{G}_s^g)$ (using equation \ref{eq:pt}) to the tangent space at $\bar{\mathcal{G}}_{s\text{ = Reference}}^g$. We choose Subject 0 as the reference.
        \vspace{0.25cm}
        
		\STATE Map all the parallelly transported vectors in the tangent space of $\bar{\mathcal{G}}_{s\text{ = Reference}}^g$ back to the manifold space using equation \ref{eq:expmap}.
	\end{algorithmic}
    \label{alg:A}
\end{algorithm}
\noindent For visualization purposes in figure \ref{fig:align}, we assume ground truth to identify clusters of a particular gesture {\em g} in {\em G}. In practice, EMG signals corresponding to different hand gestures can be classified using unsupervised methods such as {\em k}-medoids clustering. In this case, the accuracy of parallel transport is bounded by the accuracy of the unsupervised classifier. Algorithm \ref{alg:A} presents a method for aligning multivariate EMG time series from different individuals to the same neighborhood in the manifold space.
\newpage
\onecolumn
\section{Detailed results}
Here, we present detailed subject-wise results for \textsc{UCD-MyoVerse-Hand-0} and \textsc{Ninapro: Database 2-Exercise 1} datasets
.
\subsection{Results for \textsc{UCD-MyoVerse-Hand-0}}
\label{apd:res1}
The \textsc{UCD-MyoVerse-Hand-0} database consists of 30 subjects. Each subject performed 10 different hand gestures (a total of 360 trials: 10 gestures, each repeated 36 times). Therefore, the chance-level decoding accuracy is $\frac{1}{10} = 0.1$. We present the detailed subject-wise results in table \ref{tab:UCDMyoVerseSubjectwise}.
\begin{table}[htbp]
\footnotesize
\renewcommand{\arraystretch}{1.1} 
\setlength{\tabcolsep}{4pt} 
\centering
\begin{tabular}{|c|p{1.6cm}|p{1.6cm}|p{1.6cm}|} 
\hline
\multirow{2}{*}{Subject number} & \multicolumn{3}{c|}{Classification methods} \\
\cline{2-4}
& MDM & SVM {\scriptsize ($\gamma = 1$)} & {\em k}-medoids \\
\hline
0 & 0.97 & 0.99 & 0.94 \\
\hline
1 & 0.61 & 0.69 & 0.48 \\
\hline
2 & 0.62 & 0.76 & 0.61 \\
\hline
3 & 0.76 & 0.83 & 0.60 \\
\hline
4 & 0.92 & 0.93 & 0.74 \\
\hline
5 & 0.82 & 0.86 & 0.63 \\
\hline
6 & 0.94 & 0.97 & 0.88 \\
\hline
7 & 0.94 & 0.95 & 0.74 \\
\hline
8 & 0.97 & 0.97 & 0.93 \\
\hline
9 & 0.92 & 0.94 & 0.74 \\
\hline
10 & 0.84 & 0.86 & 0.62 \\
\hline
11 & 0.79 & 0.81 & 0.62 \\
\hline
12 & 0.95 & 0.97 & 0.93 \\
\hline
13 & 0.84 & 0.84 & 0.73 \\
\hline
14 & 0.82 & 0.84 & 0.64 \\
\hline
15 & 0.57 & 0.66 & 0.50 \\
\hline
16 & 0.92 & 0.93 & 0.76 \\
\hline
17 & 0.85 & 0.89 & 0.85 \\
\hline
18 & 0.53 & 0.57 & 0.35 \\
\hline
19 & 0.82 & 0.84 & 0.61 \\
\hline
20 & 0.83 & 0.87 & 0.72 \\
\hline
21 & 0.82 & 0.89 & 0.72 \\
\hline
22 & 0.93 & 0.94 & 0.70 \\
\hline
23 & 0.88 & 0.92 & 0.91 \\
\hline
24 & 0.55 & 0.55 & 0.26 \\
\hline
25 & 0.78 & 0.82 & 0.55 \\
\hline
26 & 0.83 & 0.89 & 0.83 \\
\hline
27 & 0.78 & 0.86 & 0.68 \\
\hline
28 & 0.99 & 1.0  & 0.99 \\
\hline
29 & 0.89 & 0.91 & 0.72 \\
\hline
{\bf Average} & {\bf 0.82} & {\bf 0.86} & {\bf 0.70} \\
\hline
\end{tabular}
\caption{Classification accuracy for {\em 30} subjects in  \textsc{UCD-MyoVerse-Hand-0}. Data from the first 3 sessions are used for training (first 18 repetitions of each gesture), and data from the last 3 sessions (last 18 repetitions of each gesture) are used for testing the MDM and SVM algorithms.}
\label{tab:UCDMyoVerseSubjectwise}
\end{table}

\subsection{Results for \textsc{Ninapro: Database 2-Exercise 1}}
\label{apd:res2}
The \textsc{Ninapro: Database 2-Exercise 1} consists of 40 subjects. Each subject performed 17 different hand gestures (a total of 102 trials: 17 gestures, each repeated 6 times). Therefore, the chance-level decoding accuracy is $\frac{1}{17} = 0.059$. We present the detailed subject-wise results in table \ref{tab:ninaproSubwise}.
\begin{table}[h!]
    \renewcommand{\arraystretch}{1.1}
    \setlength{\tabcolsep}{4pt}
 	\centering
 	\begin{tabular}{|c|p{1.6cm}|p{1.6cm}|p{1.6cm}|}
 		\hline
 		\multirow{2}{*}{Subject number} & \multicolumn{3}{c|}{Classification methods} \\
 		\cline{2-4}
 		& MDM & \scriptsize{SVM ($\gamma = 8$)}&  {\em k}-medoids\\
 		\hline
 		0 &  1.0& 1.0 & 0.90\\
 		\hline
 		1 & 0.94& 0.97 & 0.86\\
 		\hline
 		2 &  0.97& 0.97&  0.78\\
 		\hline
 		3 &  0.94& 0.94&  0.79\\
 		\hline
 		4 &   0.97& 1.0&  0.85\\
 		\hline
 		5 &  0.85 & 0.85&  0.79\\
 		\hline
 		6 &  0.88& 0.88&  0.69\\
 		\hline
 		7 & 0.85 & 0.91&  0.80\\
 		\hline
 		8 &  0.97& 1.0&  0.85\\
 		\hline
 		9 &  0.88& 0.91& 0.87\\
 		\hline
 		10 &  0.97& 0.91&  0.84\\
 		\hline
 		11 &  0.85& 0.91&  0.79\\
 		\hline
 		12 &  0.97& 0.94&  0.89\\
 		\hline
 		13 &  1.0& 0.97&  0.93\\
 		\hline
 		14 &  0.91& 0.97& 0.81 \\
 		\hline
 		15 &  0.91& 0.88&  0.66\\
 		\hline
 		16 &  0.97& 0.97&  0.78\\
 		\hline
 		17 &  0.88&  0.91&  0.83\\
 		\hline
 		18 &  0.91& 0.97&  0.85\\
 		\hline
 		19 &  0.88& 0.88&  0.83\\
 		\hline
 		20 &  0.88& 0.94&  0.84\\
 		\hline
 		21 &  0.91& 0.91&   0.85\\
 		\hline
 		22 &  0.88& 0.88&  0.83\\
 		\hline
 		23 &  0.91& 0.94&  0.78\\
 		\hline
 		24 &  0.91& 0.94& 0.84 \\
 		\hline
 		25 &  0.97& 0.97&  0.85\\
 		\hline
 		26 &  0.97& 0.94&  0.87\\
 		\hline
 		27 &  0.91& 0.94&  0.87\\
 		\hline
 		28 &   0.91& 0.88&  0.75\\
 		\hline
 		29 &  0.82& 0.91&  0.76\\
 		\hline
 		30 &  0.88& 0.88&  0.68\\
 		\hline
 		31 &  0.91& 0.88&  0.84\\
 		\hline
 		32 &  0.94& 0.97&  0.85\\
 		\hline
 		33 &  0.76& 0.82&  0.73\\
 		\hline
 		34 &  0.97& 0.97&  0.84\\
 		\hline
 		35 &  0.97& 0.88&  0.80\\
 		\hline
 		36 &  0.94& 1.0&  0.89\\
 		\hline
 		37 &  0.97& 0.97&  0.96\\
 		\hline
 		38 &  0.97& 0.94& 0.84 \\
 		\hline
 		39 &  0.97& 0.97&  0.86\\
 		\hline
 		{\bf Average} &  \bf{0.92}& \bf{0.93}&  \bf{0.82}\\
 		\hline
 	\end{tabular}
 \caption{Classification accuracy for 40 subjects in \textsc{Ninapro: Database 2-Exercise 1}. Following the works in  \citeauthor{9463745} and \citeauthor{rahimian2021temgnet}, for each gesture, we use repetitions {\em 1, 3, 4}, and {\em 6} for training and repetitions {\em 2} and {\em 5} for testing of  MDM and SVM algorithms.}\label{tab:ninaproSubwise}
 \end{table}
\end{document}